\documentclass[aps,pra,twocolumn,amsmath,amssymb]{revtex4-2}

\usepackage{graphicx}
\usepackage{color}
\usepackage[hidelinks]{hyperref}

\hypersetup{colorlinks=true, linkcolor=black, citecolor=black, urlcolor=blue}

\begin{document}
\author{Karol Gietka}
\email[]{karol.gietka@uibk.ac.at}

\affiliation{Institut f\"ur Theoretische Physik, Universit\"at Innsbruck, A-6020 Innsbruck, Austria} 

\title{Squeezing of the quantum electromagnetic vacuum}

\begin{abstract}
It is commonly agreed that the electromagnetic vacuum is not empty but filled with virtual photons. This leads to effects like Lamb shift and spontaneous emission. Here we argue that if the vacuum has virtual photons it might mean that it is very weakly squeezed and therefore the electromagnetic field is not in its ground state (vacuum) but in an excited dark state. We suggest a stringent test relying on measuring various properties of the electromagnetic field to exclude this yet-untested squeezing hypothesis. This could be done by measuring the number of photons as a function of frequency and comparing it with the spectrum of electric (or magnetic) field fluctuations. If such squeezing exists, it might shed new light on cosmological phase transitions and give complementary information to the observed microwave background radiation as well as be a possible candidate for dark energy.
\end{abstract}
\date{\today}
\maketitle

\section{Introduction}

Squeezing~\cite{gardiner2004quantum,Zubairy_2005_squeezing} is predicted to play a key role in various quantum technologies, in particular, in quantum-enhanced measurements~\cite{quantummetrology2004}. It relies on redistributing quantum uncertainties between two non-commuting observables. The primary example is the squeezing of light~\cite{squeezing1983walls}, where the uncertainties are redistributed between the strength of electric and magnetic fields with respect to a coherent state where the uncertainties are equal. The first squeezed light was observed by Slusher \emph{et al.} in 1985~\cite{slusher1985squeezed}. Since that time, a number of experiments have reported on generating squeezed light~\cite{PhysRevLett1986wu,machida1987,wodkiewicz1987} using various platforms and characterized with larger and larger squeezing~\cite{Andersen_2016}. The current record for direct measurement of squeezing is the measurement of 15 dB squeezed vacuum state of light~\cite{PhysRevLett.117.110801}. In this manuscript, however, we suggest that instead of measuring even more squeezed states of light one should focus on measuring the tiniest amount of squeezing of the electromagnetic field.

The quantum electromagnetic vacuum is the lowest energy state of the quantized electromagnetic field~\cite{cao2004conceptual}
\begin{align}
    \hat H = \sum \hat a^\dagger_{\omega} \hat a_{\omega},
\end{align}
where the sum goes over all the frequencies $\omega$, directions, and polarizations. It is commonly agreed that the electromagnetic vacuum is not really empty but filled with virtual particles which give rise to vacuum fluctuations~\cite{loudon2000quantum}. These fluctuations in turn affect the energy levels of atoms leading to effects such as spontaneous emission~\cite{scully1999quantum}, Casimir force~\cite{casimir1948attraction,Lamoreaux_2005}, and Lamb shift~\cite{lambshif1947,1947bethe,science2008lamb}. The intuitive picture provided to explain the virtual particles is based on the Heisenberg uncertainty principle
\begin{align}
    \Delta E \Delta t \geq \frac{\hbar}{2},
\end{align}
where $\Delta E$ and $\Delta t$ are the uncertainties of energy and time, respectively. It is sometimes argued that because the lifetime of the virtual particle is very short, it can, in certain sense, \emph{borrow} the energy from the vacuum and pop into existence~\cite{davies1982accidental}. Such an explanation is, however, \emph{ad hoc} and is often criticized~\cite{king2001quest}. In particular, there is no time operator in quantum mechanics, therefore, the energy and time do not satisfy a canonical commutation relation~\cite{busch2008time}. An alternative explanation to the virtual particles and the fluctuations of the electromagnetic field that does not rely on the energy-time uncertainty relation is the squeezing of the electromagnetic vacuum. We suggest in this manuscript that the state of the electromagnetic field might not be in its ground state but in an excited squeezed dark state~\cite{lambropoulos2007fundamentals} which simply contains virtual particles by definition. In order to understand why this might be the case, we consider a cavity quantum electrodynamics setup~\cite{helmut2013rmp} with only one mode of the electromagnetic field coupled to a two-level atom. We show that the vacuum state can be squeezed by exploiting light-matter interactions. If these interactions are weak and subsequently turned off (non-adiabatically), the state of the system can remain in an excited squeezed dark state. We suggest ways in which squeezing of the electromagnetic vacuum can be measured to exclude this yet-untested hypothesis. Finally, we attempt to give a plausible explanation of why the state of the electromagnetic vacuum might be squeezed in the first place.

\begin{figure*}[htb!]
    \centering
    \includegraphics[width=\textwidth]{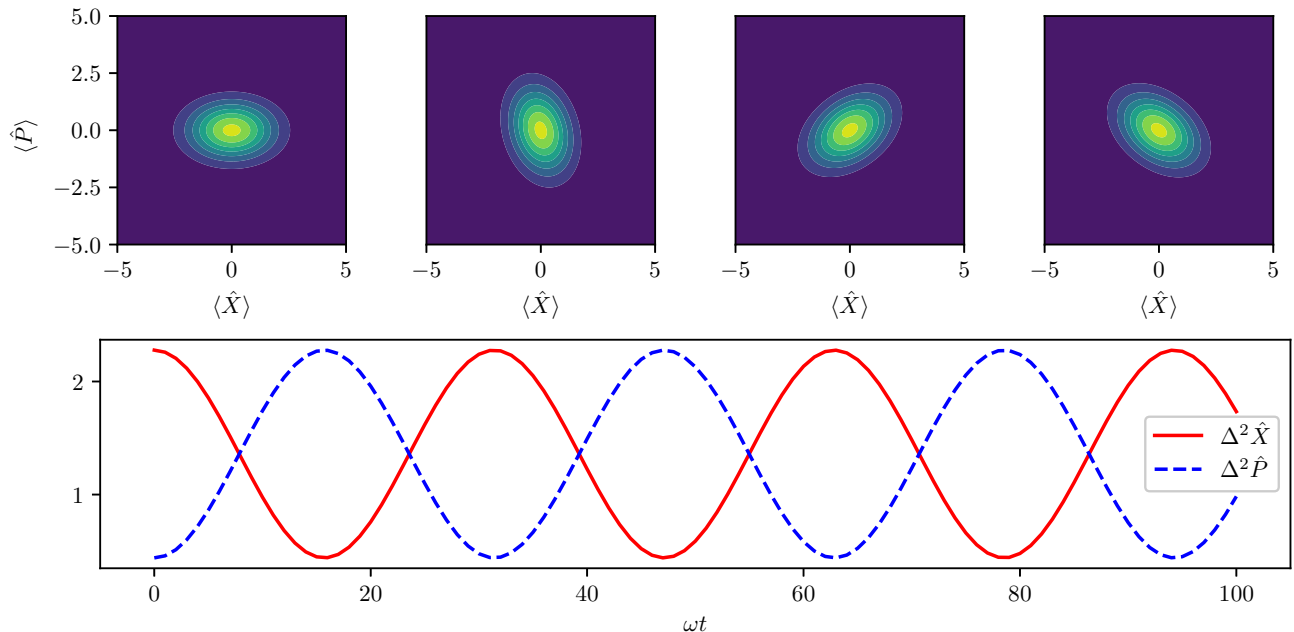}   
    \caption{The rotating squeezed single-mode electromagnetic vacuum as a function of time. The top row shows the Husimi function of the squeezed electromagnetic vacuum at various times. The bottom row shows the time-dependent fluctuations of the quadratures $\hat X$ (related to the electric field) and $\hat P$ (related to the magnetic field). Although $\langle \hat n \rangle >0$, these excitations might be almost impossible to measure if the squeezing is extremely weak (virtual photons). For the sake of illustration, the amount of squeezing is exaggerated. The quantum electromagnetic vacuum might consists of infinitely many squeezed single-mode vacua each rotating with its own frequency at every point in space.}
    \label{fig:fig1}
\end{figure*}

\section{Squeezing of a single-mode electromagnetic vacuum}

Let us consider an atom interacting with a single mode of a cavity in which it resides. The Hamiltonian of such a system is given by the paradigmatic quantum Rabi model
\begin{align}
    \hat H = \omega \hat a^\dagger \hat a + \frac{\Omega}{2}\hat \sigma_z + \frac{g}{2}\left(\hat a + \hat a^\dagger \right) \hat \sigma_x,
\end{align}
where $\hat a$ describes the electromagnetic field with frequency $\omega$, the Pauli matrices describe the two-level atom with frequency $\Omega$, and $g$ is the coupling between the atom and the electromagnetic field mode. If the frequency of the atom $\Omega$ is larger than the frequency of the field $\omega$ and the coupling strength is not too strong $1-g^2/g_c^2>(\omega/\Omega)^{2/3}$~\cite{gietka2022harnessicncom}, the Schrieffer-Wolff transformation~\cite{PhysRev.149.491} can be used to eliminate the two-level atom from the Hamiltonian leading to an effective description~\cite{puebla2015universalqrm,gietka2022speedup} (note that squeezing can be also obtained for $\omega \geq \Omega$)
\begin{align}
    \hat H_\mathrm{eff} \approx \omega \hat a^\dagger \hat a - \frac{g^2}{4\Omega}\left(\hat a + \hat a^\dagger \right)^2,
\end{align}
which can be rewritten using the abstract position and momentum operators $\hat x = (\hat a + \hat a^\dagger)/\sqrt{2\omega}$ and $\hat p = \sqrt{\omega}(\hat a - \hat a^\dagger)/\sqrt{-2}$ in the form of a harmonic oscillator 
\begin{align}
     \hat H_\mathrm{eff} = \frac{\hat p^2}{2} + \frac{\omega^2}{2}\left(1-\frac{g^2}{g_c^2}\right)\hat x^2.
\end{align}
The above Hamiltonian describes an abstract harmonic oscillator with unit mass and a frequency $\omega \sqrt{1-g^2/g_c^2}$. Once $ g > 0$ the abstract harmonic oscillator ground state is squeezed with respect to the physical harmonic oscillator ($\hat a^\dagger \hat a$) ground state. In terms of the physical harmonic oscillator operators, the ground state wavefunction (also known as a polariton~\cite{polariton2012,2017polariton2,2017polariton3}) can be described as
\begin{align}\label{eq:fullsv}
    |\psi_0 \rangle = \exp\bigg\{\frac{1}{2}\left(\xi^*\hat a^2-\xi\hat a^{\dagger2}\right)\bigg\}|0\rangle,
\end{align}
where $\xi = \frac{1}{4} \ln\{1-g^2/g_c^2\}$ is the squeezing parameter and $|0\rangle$ is the ground state of the physical harmonic oscillator. The number of photons of such a state can be easily calculated to be
\begin{align}
    \langle \hat n \rangle = \sinh^2\xi \geq 0.
\end{align}
Although the number of photons is greater than 0 for a non-zero $\xi$, these photons do not correspond to real photons (radiation) but to the virtual photons~\cite{carusotto2005,virtual2009nature}. This happens because for a non-zero coupling, the ground state of the electromagnetic field is squeezed but the ground state cannot lose the energy (otherwise it would constitute a \emph{perpetuum mobile}). If the coupling is suddenly turned off, these virtual photons can become real photons because the squeezed vacuum is no longer the ground state of the system. However, if the squeezing is very weak, the detection of such squeezing might be quite difficult~\cite{difficultysqueezing1985}. In order to understand it, let us look at the squeezed single-mode vacuum state from Eq.~\eqref{eq:fullsv} expressed in the fock basis
\begin{align}
    |\psi_0 \rangle = \frac{1}{\sqrt{\cosh{|\xi|}}}\sum_{n=0}^\infty (e^{-i\xi/|\xi| \tanh |\xi|})^n \frac{\sqrt{(2n)!}}{2^n n!}|2n\rangle.
\end{align}
If the squeezing is very weak, the above equation can be approximated as
\begin{align}
     |\psi_0 \rangle \approx \frac{1}{\sqrt{\cosh{|\xi|}}}|0\rangle + (e^{-i\xi/|\xi| \tanh |\xi|})\sqrt{\frac{\cosh{|\xi|}-1}{{\cosh{|\xi|}}}} |2\rangle.
\end{align}
For weak squeezing $|\xi|\ll 1$ it might be virtually impossible to detect the photons, thus measuring no photons for a given time only pushes the limit for $\xi$ towards lower values (not measuring photons does not project the system to a state with no photons). The photons would have to be measured in total darkness so essentially the dark counts would be measured in the experiment. Also measuring these photons itself does not necessarily indicate squeezing; there might be simply photons traveling around and hitting the detectors. Therefore measuring the number of photons only to characterize the weak squeezing would not lead to any meaningful conclusions. However, there is one more effect that such a state will lead to. Since a squeezed vacuum is not the eigenstate of the electromagnetic field, it will rotate in the phase space leading to time-dependent fluctuations of the electric and magnetic field. This is illustrated in Fig. 1, where we show the Husimi function of the squeezed vacuum field at various times and the corresponding variances of the quadrature operators. 

\section{Squeezing of the quantum electromagnetic vacuum}

\begin{figure}[htb!]
    \centering
    \includegraphics[width=0.5\textwidth]{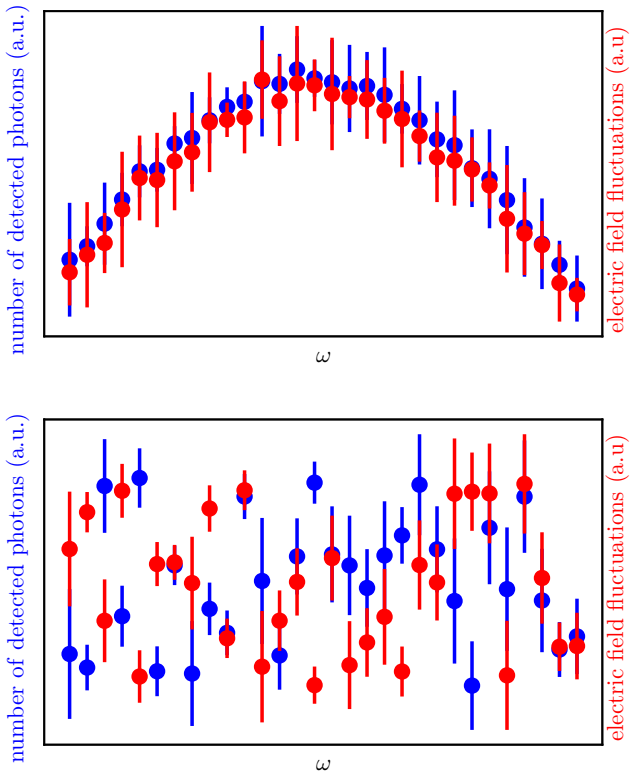}   
    \caption{Illustration showing the possible conclusive measurement outcomes. The blue color represents the number of photons and the red color represents the fluctuations of the field. (top panel) If the measurement of the number of photons as a function of frequency shows similar behavior to the fluctuations of the electric (magnetic) field, it might be a confirmation that the state of the quantum electromagnetic vacuum is weakly squeezed. (bottom panel) If the number of photons as a function of frequency does not correspond to the fluctuations of the electric (magnetic) field, the two phenomena are not related to each other and the state of the electromagnetic vacuum is not squeezed. }
    \label{fig:fig2}
\end{figure}

Once we understand how rotating single-mode squeezed vacuum leads to time-dependent fluctuations of the electromagnetic field, we can proceed to propose a measurement that will be capable of quantifying the amount of squeezing for each mode of the electromagnetic field. In particular, it should show if the electromagnetic vacuum is squeezed at all. Measuring all the modes at once might be too complicated to analyze. Ideally, one should measure how each mode of the electric (or magnetic) field fluctuates in time. Then the amplitude of the fluctuations will be related to the squeezing parameter $\xi(\omega)$ of that mode (we choose the electric field because it is stronger than the magnetic field)
\begin{align}\label{eq:fluc}
\begin{split}
    \Delta^2 \hat E(\omega) \sim \Delta^2 \hat X(\omega) = &\frac{1}{2}\exp(2\xi)\cos^2(\omega t) \\&+ \frac{1}{2}\exp(-2 \xi)\sin^2(\omega t).
    \end{split}
\end{align}
However, a direct measurement of the electric field fluctuations~\cite{science2015direct,2017direc2} would not reveal squeezing as the fluctuations would average out over time and it would only indicate the amplitude of the possible squeezing. Although the uncertainty might be larger than predicted by the Heisenberg uncertainty principle for electric and magnetic field~\cite{lemaitre1933} it would be very hard to show it because the fluctuations would have to be measured in a single point or in a precisely known volume. Only an electric field correlation measurement for various space-time coordinates would reveal that the electric field fluctuations behave as in Eq.~\eqref{eq:fluc} (for a time-independent groundstate there would be no correlations). Such a measurement was reported in Ref.~\cite{electricfield2019faist}, where the authors measured the fluctuations in the terahertz frequency range. This was achieved by using electro-optic detection~\cite{1995electroptic} in a nonlinear crystal placed in a cryogenic environment. In particular, in Ref.~\cite{electricfield2019faist} it was shown experimentally that the electro-optic field correlation measurement on a vacuum state is non-zero. Furthermore, in a recent experiment~\cite{2022quantumvacuumfieldcorrelations}, it was shown that the quantum vacuum field fluctuations are correlated outside the light cone---a signal that a uniformly squeezed vacuum would generate. These two results alone, however, would not confirm yet that the quantum electromagnetic vacuum is squeezed. Once the amplitude and correlations are measured for the large part of the spectrum, these measurements should be subsequently compared with the results from photon detection. If all these results show no correlations it would exclude the possibility of squeezing of the quantum electromagnetic vacuum. In other words, measuring single photons (dark counts) from the alleged squeezed vacuum for a sufficiently long time will eventually lead to enough collected data to create a histogram of photons as a function of frequency. If the electric field fluctuations and its amplitude as a function of frequency resemble the created histogram, it seems that squeezing of the quantum electromagnetic vacuum is the simplest possible explanation for all the observed phenomena. In principle, one might come up with even more tests based on measuring the fluctuations of the magnetic field. However, as quantum mechanically electric and magnetic fields are conjugate variables satisfying canonical commutation relations, the previously described tests seem sufficient to test the squeezing hypothesis. The possible measurement outcomes are schematically illustrated in Fig.~\ref{fig:fig2}

\section{plausible explanation of vacuum squeezing}

One may wonder why the electromagnetic vacuum might be squeezed in the first place, therefore we give now a plausible explanation. According to the cosmology theory, a long time ago, approximately until around 370,000 years after the Big Bang~\cite{tabashi2018}, the matter was strongly interacting with the electromagnetic field. Due to the Thomson scattering~\cite{thomson1988scsttering,thomson2002conduction} by free electrons, the mean distance that a photon could travel before interacting with an electron was very short similar to the mean distance that a photon can travel in a cavity before interacting with an atom if the light-matter coupling $g$ is strong enough. This suggests that in analogy to the squeezing of the electromagnetic field in the cavity, the field ground state of the coupled light-matter system in the early Universe might have been squeezed as well. The subsequent recombination due to cooling caused by expanding Universe reduced the number of electrons (they started forming Hydrogen atoms with protons), and thus \emph{released} the photons. In other words, the cooling of the Universe might have effectively non adiabatically decreased the coupling between light and matter. The real photons started to roam around the Universe and are observable today through cosmic microwave background radiation~\cite{penzias1965measurement}. However, if the electromagnetic field was indeed squeezed, this squeezing might have survived until today giving rise to a number of physical phenomena. If this is the case, such squeezing could contain the missing information about the early stage of the Universe that the microwave background radiation does not have because these photons were not there at that time. Moreover, if the state of the electromagnetic field is excited, this excess energy with respect to the true vacuum might be a candidate for dark energy~\cite{darkenergy2003rmp,darkenergy2006} because the electromagnetic vacuum should be squeezed uniformly across the entire space.

In order to understand why weak squeezing of the electromagnetic vacuum might lead to measurable effects let us consider a phenomenon from special relativity. If a small mass (of the order of a gram) is moving with a small velocity with respect to the speed of light, its change of mass will be negligible and barely measurable. If a big mass (a massive star) starts to move with a small velocity with respect to the speed of light its absolute change of mass will be large and capable of affecting other phenomena, in particular, local gravitational field. For a single-mode electromagnetic vacuum, very weak squeezing will be negligible and barely measurable as well. However, as the electromagnetic field consists of infinitely many modes at every point in space (and two polarizations in each of the three dimensions), the effect of very weak squeezing might be affecting other phenomena similarly as in the example with a massive star (which is essentially constructed of many small masses).

\section{conclusions}

In conclusion, we have presented an alternative explanation for the fluctuations of the electromagnetic field and the virtual particles. This explanation assumes that the electromagnetic vacuum is not in its ground state but in a weakly excited squeezed dark state which contains virtual particles by definition. We proposed an experiment able to confirm or refute whether the electromagnetic vacuum is squeezed at all testing thus the squeezing hypothesis, and we provided a plausible explanation of why the electromagnetic vacuum might be squeezed in the first place. To this end, we used a simple example known from the cavity quantum electrodynamics and established an analogy with the early stage of the Universe. Finally, we suggested that such weak squeezing of the electromagnetic vacuum might be a candidate for dark energy. %Therefore, until evidence is provided against the squeezing hypothesis, the squeezing of the quantum electromagnetic vacuum should be considered a valid possibility.

\section{acknowledgements}

Simulations were performed using the open-source \textsc{QuantumOptics.jl}~\cite{kramer2018quantumoptics} framework in \textsc{Julia}. This work was supported by the Lise-Meitner Fellowship M3304-N of the Austrian Science Fund (FWF).

\end{document}